# Robustness of MIMO-OFDM Schemes for Future Digital TV to Carrier Frequency Offset

Youssef Nasser, Jean-François Hélard, Matthieu Crussière

*Institute of Electronics and Telecommunications of Rennes, UMR CNRS 6164, Rennes, France*

*20 Avenue des Buttes des Coesmes, 35043 Rennes cedex, France*

Email : youssef.nasser@insa-rennes.fr

**Abstract-** This paper investigates the impact of carrier frequency offset (CFO) on the performance of different MIMO-OFDM schemes with high spectral efficiency for next generation of terrestrial digital TV. We show that all studied MIMO-OFDM schemes are sensitive to CFO when it is greater than 1% of inter-carrier spacing. We show also that the Alamouti scheme is the most sensitive MIMO scheme to CFO.

***Keywords-*** *Modulation & Multiplexing (MIMO-OFDM), Signal Processing for Transmission (Carrier Frequency Offset).*

## I. INTRODUCTION

Digital video broadcast is the technology driving fixed, portable and mobile TV. Since its inauguration in 1993, digital video broadcast (DVB) project for terrestrial (DVB-T) transmission has fully responded to the objectives of its designers, delivering wireless digital TV services in almost every continent [1]. The main concern of many researchers is to support transmission at higher data rates with minimum error probability. In 2006, the DVB forum launched a study mission to investigate what technologies might be considered for a future DVB-T2 standard. It is expected that a multiple input multiple output (MIMO) system combined with orthogonal frequency division multiplexing (OFDM) should take place for that target. However, it is well known that OFDM systems suffer considerably from carrier frequency offset (CFO) between transmitter and receiver since CFO includes inter carrier interference (ICI) at the receiving side [2].

This work is carried out within the framework of the European project '*Broadcast for the 21$^{st}$ Century*' (B21C) which constitutes a contribution task force to the reflections engaged by the DVB forum. The main contribution of this work is twofold. First, a generalized framework is proposed for modelling the effect of CFO on MIMO-OFDM systems. Therefore, we analyze the robustness of different MIMO-OFDM schemes to CFO using a sub-optimal iterative receiver. This analysis should give a global view on the best suitable MIMO-OFDM scheme with respect to CFO.

The paper is organised as follows. Section 2 describes the transmission system model. In section 3, we present the different MIMO schemes considered in this paper. Section 4 gives some simulation results. Conclusions are drawn in section 5.

## II. SYSTEM MODEL

We consider in this paper the downlink communication with two transmit antennas ($M_T$=2) at the base station and two receiving antennas ($M_R$=2) at the terminal. Figure 1 depicts the transmitter modules. Information bits $b_k$ are first channel encoded with a convolutional encoder of coding rate $R$. The encoded, interleaved bits are then fed to a quadrature amplitude modulation (QAM) module which assigns $B$ bits for each of the complex constellation points. Therefore, each group **s**=[$s_1$,…,$s_Q$] of $Q$ complex symbols is encoded through a space time (ST) block code (STBC) encoder and transmitted during $T$ symbol durations according to the chosen ST scheme. The ST coding rate is then defined by $L=Q/T$. With $M_T$ transmitting antennas, the output of the ST encoder is an ($M_T,T$) matrix **X**=[$x_{i,t}$] where $x_{i,t}$ ($i$=1,…,$M_T$; $t$=1,…,$T$) is a function of the input symbols $s_q$ ($q$=1,…,$Q$) depending on STBC encoder type. The resulting symbols are then fed to OFDM modulator of $N$ subcarriers.

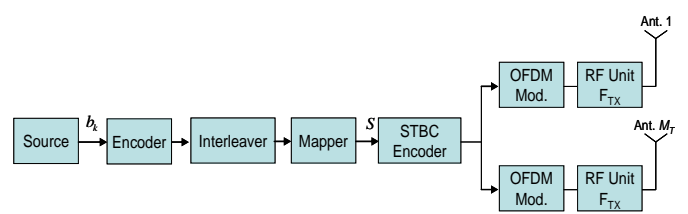

Figure 1- Block diagram of the transmitter

After D/A conversion, the signal is transposed to the transmitter carrier frequency $F_{TX}$ by the RF unit, and transmitted through the channel. At the receiver (Figure 2), it is transposed to base band with the receiver carrier frequency $F_{RX}$ and sampled at sampling frequency $F_s=1/T_s$. In this work, we assume equal carrier frequencies $F_{TX}$ for all transmitting antennas and equal carrier frequencies $F_{RX}$ for all receiving antennas. The carrier frequency offset is therefore given by $\Delta F = F_{RX} - F_{TX}$. After OFDM demodulation, the signal received by the $j^{th}$ antenna at each time sample $t$ on the $n^{th}$ subcarrier could be written as:

$$y_j[n,t] = \frac{1}{\sqrt{M_T}} \sum_{m=0}^{M_T-1} \sum_{p=0}^{N-1} x_i[p,t] h_{j,i}[p] \varphi(n,p) + w_j[n,t] \quad (1)$$

zero mean and $N_0/2$ variance. $\phi(n,p)$ is a function of the CFO, given by:

$$\phi(n,p) = e^{j\pi \frac{N-1}{N}(N\Delta F T_s + (n-p))} \frac{1}{N} \frac{\sin(\pi(N\Delta F T_s + (n-p)))}{\sin(\pi(N\Delta F T_s + (n-p))/N)} \quad (2)$$

The signal received by the $M_R$ antennas on sub-carrier $n$ are gathered in a matrix $\mathbf{Y}[n]$ of dimension $(M_R, T)$. It can be deduced from (1) by:

$$\mathbf{Y}[n] = \varphi(n,n)\mathbf{H}[n]\mathbf{X}[n] + \sum_{\substack{p=1 \\ p \neq n}}^{N} \varphi(n,p)\mathbf{H}[p]\mathbf{X}[p] + \mathbf{W}[n]$$

$$= \mathbf{H_{eq}}[n]\mathbf{X}[n] + \sum_{\substack{p=1 \\ p \neq n}}^{N} \varphi(n,p)\mathbf{H}[p]\mathbf{X}[p] + \mathbf{W}[n] \quad (3)$$

In (3), the first term represents useful signal, the second term indicates the ICI and the last one is the AWGN. $\phi(n,n)$ can be seen as a phase rotation and an amplitude distortion of the useful signal due to CFO. The ICI could be seen as an additive noise to the useful signal. It will be neglected in the equalization process. $\mathbf{H}[n]$ is a $(M_R, M_T)$ matrix whose components are the channel coefficients $h_{j,i}[n]$, $X[n]$ is a $(M_T, T)$ matrix whose components are the transmitted symbols on the $M_T$ antennas during $T$ OFDM symbols on the $n^{th}$ subcarrier and $W[n]$ is the AWGN.

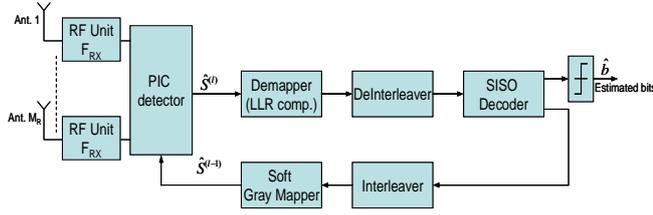

Figure 2- Iterative receiver structure with parallel interference cancellation detector

Let us now describe the transmission link with a general model independently of the ST coding scheme. We separate the real and imaginary parts of the complex symbols input vector $\mathbf{s}$ $\{s_q: q=1,\ldots,Q\}$, of the outputs $\mathbf{X}$ of the double layer ST encoder as well as those of the channel matrix $\mathbf{H}$, and the received signal $\mathbf{Y}$. Let $s_{q,R}$ and $s_{q,I}$ be the real and imaginary parts of $s_q$. The main parameters of the double code are given by its dispersion matrices $\mathbf{U_q}$ and $\mathbf{V_q}$ corresponding (not equal) to the real and imaginary parts of $\mathbf{X}$ respectively. With these notations, $\mathbf{X}$ is given by:

$$\mathbf{X} = \sum_{q=1}^{Q} \left( s_{q,\Re} \mathbf{U_q} + j s_{q,\Im} \mathbf{V_q} \right) \quad (4)$$

We separate the real and imaginary parts of $\mathbf{S}$, $\mathbf{Y}$ and $\mathbf{X}$ and stack them row-wise in vectors of dimensions $(2Q,1)$, $(2M_RT,1)$ and $(2M_TT,1)$ respectively. We obtain:

where $h_{j,i}[p]$ is the frequency channel coefficient on the $p^{th}$ subcarrier assumed constant during $T$ OFDM symbols, $W_j[n]$ is the additive white Gaussian noise (AWGN) with

$$\mathbf{s} = \left[ s_{1,\Re}, s_{1,\Im}, \ldots, s_{Q,\Re}, s_{Q,\Im} \right]^{tr}$$

$$\mathbf{y} = \left[ y_{1,\Re}, y_{1,\Im}, \ldots, y_{T,\Re}, y_{T,\Im}, \ldots, y_{M_RT,\Re}, y_{M_RT,\Im} \right]^{tr} \quad (5)$$

$$\mathbf{x} = \left[ x_{(1,1),\Re}, x_{(1,1),\Im}, \ldots, x_{(2M_T,T),\Re}, x_{(2M_T,T),\Im} \right]^{tr}$$

where $tr$ holds for matrix transpose.

Since, we use linear ST coding, the vector $\mathbf{x}$ can be written as:

$$\mathbf{x} = \mathbf{F} . \mathbf{s} \quad (6)$$

where $\mathbf{F}$ has the dimensions $(2M_TT, 2Q)$ and is obtained through the dispersion matrices of the real and imaginary parts of $\mathbf{X}$. We obtain the vector $\mathbf{y}[n]$ given by:

$$\mathbf{y}[n] = \mathbf{G}[n]\mathbf{F}\mathbf{s}[n] + \sum_{\substack{p=1 \\ p \neq n}}^{N} \varphi(n,p)\mathbf{G}[p]\mathbf{F}\mathbf{s}[p] + \mathbf{w}[n]$$

$$= \mathbf{G_{eq}}[n]\mathbf{s}[n] + \sum_{\substack{p=1 \\ p \neq n}}^{N} \varphi(n,p)\mathbf{G_{eq}}[p]\mathbf{s}[p] + \mathbf{w}[n] \quad (7)$$

with $\mathbf{G_{eq}}[n] = \mathbf{G}[n].\mathbf{F}$

where $\mathbf{G}[n]$ is composed of blocks $\mathbf{G_{j,i}}$ ($j=1,\ldots,M_R$; $i=1,\ldots,M_T$) each having $(2T,2T)$ elements [3] given by:

$$G_{i,j} = \begin{pmatrix} h_{(i,j),\Re} & -h_{(i,j),\Im} & 0 & & \cdots & & 0 \\ h_{(i,j),\Im} & h_{(i,j),\Re} & 0 & & \cdots & & 0 \\ 0 & 0 & h_{(i,j),\Re} & -h_{(i,j),\Im} & 0 & \cdots & 0 \\ 0 & 0 & h_{(i,j),\Im} & h_{(i,j),\Re} & 0 & \cdots & 0 \\ 0 & \cdots & 0 & \ddots & 0 & 0 \\ 0 & \cdots & 0 & \ddots & 0 & 0 \\ 0 & \cdots & & & 0 & h_{(i,j),\Re} & -h_{(i,j),\Im} \\ 0 & \cdots & & & 0 & h_{(i,j),\Im} & h_{(i,j),\Re} \end{pmatrix}_{(2T,2T)} \quad (8)$$

In this work, we use an iterative receiver for non-orthogonal (NO) schemes where the ST detector and channel decoder exchange extrinsic information in an iterative way until the algorithm converges. The iterative detection and decoding exploits the error correction capabilities of the channel code to provide improved performance. The estimated symbols $\hat{\mathbf{s}}^{(1)}$ at the first iteration are obtained via minimum mean square error (MMSE) filtering as:

$$\hat{s}_u^{(1)}[n] = \mathbf{g_u^{tr}}[n] \left( \mathbf{G_{eq}}[n].\mathbf{G_{eq}^{tr}}[n] + \sigma_w^2 \mathbf{I} \right)^{-1} \mathbf{y}[n] \quad (9)$$

where $\mathbf{g_u^{tr}}[n]$ of dimension $(1, 2M_RT)$ is the $u^{th}$ column of $\mathbf{G_{eq}}$ ($1 \leq u \leq 2Q$). $\hat{s}_u^{(1)}$ is the estimation of the real part ($u$ odd) or imaginary part (*u* even) of $s_q$ ($1 \leq q \leq Q$) at the first iteration. At each iteration, the demapper provides soft information about transmitted coded bits. The soft information is represented by log likelihood ratios (LLR). After de-interleaving, it is fed to the outer decoder which computes the '*a posteriori*' extrinsic information of the coded bits. After interleaving, this extrinsic information will be used by the soft mapper to produce estimation of transmitted QAM symbols. From the second iteration (*l*>1), we perform parallel interference cancellation (PIC) followed by a simple inverse filtering:

$$\hat{\mathbf{y}}^{(l)}[n] = \mathbf{y}[n] - \mathbf{G}_{eq,u}[n]\tilde{\mathbf{s}}_u^{(l-1)}[n]$$
$$\hat{s}_u^{(l)}[n] = \frac{1}{\mathbf{g}_u^{tr}[n]\mathbf{g}_u[n]} \mathbf{g}_u^{tr}[n]\hat{\mathbf{y}}^{(l)}[n] \quad (10)$$

where $\mathbf{G}_{eq,u}[n]$ of dimension $(2M_RT, 2Q-1)$ is the matrix $\mathbf{G}_{eq}[n]$ with its $u^{th}$ column removed, $\tilde{s}_u^{(l-1)}$ of dimension $(2Q-1, 1)$ is the vector $\tilde{s}^{(l-1)}[n]$ estimated at the previous iteration by the soft mapper with its $u^{th}$ entry removed. The exchange of information between detector and decoder runs until the process converges.

### III. CONSIDERED ST CODING SCHEMES

First, we consider the simplest orthogonal ST coding scheme proposed by Alamouti [4] as a reference of comparison. Since $M_T=2$, we have $Q=T=2$ and the ST coding rate $L=1$. This code is given by the matrix:

$$X = \begin{bmatrix} s_1 & s_2 \\ -s_2^* & s_1^* \end{bmatrix} \quad (11)$$

For non-orthogonal schemes, we consider in this work the well-known multiplexing scheme i.e. the V-BLAST [5]. VBLAST is designed to maximize the rate by transmitting symbols sequentially on different antennas. Its coding scheme is given for $T=1$, $Q=2$ and $L=2$ by:

$$X = \begin{bmatrix} s_1 & s_2 \end{bmatrix}^{tr} \quad (12)$$

We also consider the LD code proposed by Hassibi [6] for which we have $Q=4$, $T=2$ and $L=2$. It is designed to maximize the mutual information between transmitter and receiver. It is defined by:

$$X = \frac{1}{\sqrt{2}} \begin{bmatrix} s_1 + s_3 & s_2 - s_4 \\ s_2 + s_4 & s_1 - s_3 \end{bmatrix} \quad (13)$$

Finally, we consider the optimized Golden code [7] denoted hereafter by GC. The Golden code is designed to maximize the rate such that the diversity gain is preserved for an increased signal constellation size. It is defined for $Q=4$, $T=2$ and $L=2$ by:

$$X = \frac{1}{\sqrt{5}} \begin{bmatrix} \beta(s_1 + \theta s_2) & \beta(s_3 + \theta s_4) \\ \mu\bar{\beta}(s_3 + \bar{\theta} s_4) & \bar{\beta}(s_1 + \bar{\theta} s_2) \end{bmatrix} \quad (14)$$

where $\theta = \frac{1+\sqrt{5}}{2}$, $\bar{\theta} = 1-\theta$, $\beta = 1+j(1-\theta)$, $\bar{\beta} = 1+j(1-\bar{\theta})$, $\mu = j$ and $j = \sqrt{-1}$.

### IV. SIMULATION RESULTS

In this section, we present a comparative study of the four MIMO coding schemes.. The performance comparison is made for a frequency non selective channel with independent Gaussian distributed coefficients. It is computed in terms of bit error rate (BER) versus Eb/N0 ratio for different values of CFO expressed in terms of inter-carrier spacing $1/NT_s$. The simulations parameters are chosen from those of DVB-T as shown in **Table** 1.

Table 1- Simulations Parameters

| Number of subcarriers | 2K mode |
|---|---|
| Rate $R_c$ of convolutional code | 1/2, 3/4 |
| Polynomial code generator | (133,171)$_o$ |
| Channel estimation | perfect |
| Constellation | 64-QAM, 256-QAM |
| Spectral Efficiency | 6 [b/s/Hz] |

Figure 3 shows that the sensitivity of Alamouti scheme to CFO for a spectral efficiency η= 6 [b/s/Hz] becomes noticeable for a CFO such that $N\Delta FT_s \geq 1\%$ i.e. $\Delta F \geq 0.01/NT_s$ (equivalent to 5ppm). Figure 4 (respectively Figure 5) gives the Eb/N0 required to reach a BER=10$^{-4}$ (BER=10$^{-3}$) for a spectral efficiency η= 2 [b/s/Hz] (η= 6 [b/s/Hz]) and the different MIMO schemes. These figures show that for low spectral efficiency, Alamouti scheme outperforms other schemes. However, Golden code offers the best performance for high spectral efficiency.

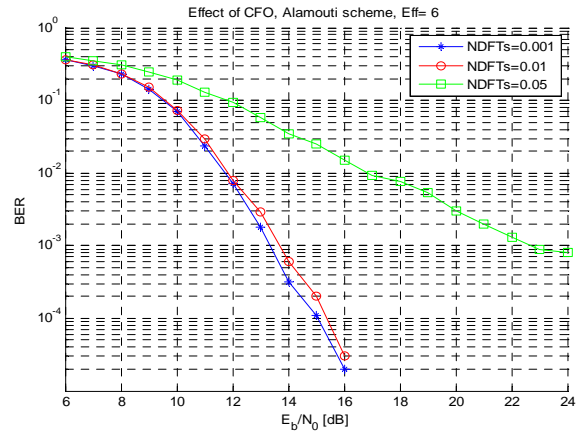

Figure 3- Effect of CFO, Alamouti scheme, Spectral efficiency η=6 [b/s/Hz] (256-QAM, R=3/4).

Moreover, Figure 5 shows that Alamouti scheme presents the worst results when η increases. Indeed, the Eb/N0 required to obtain a BER=$10^{-3}$ is about 22.8dB for $N\Delta FT_s$=0.05 (equivalent to 25ppm only for N=2048) where it is only about 13.7dB for $N\Delta FT_s$=0.01. This is due to the orthogonality loss of Alamouti scheme for higher constellation size. As a conclusion, the choice of a given MIMO-OFDM scheme is not confident for all spectral efficiencies when it is based on CFO. That is, for the second generation of digital TV transmission, other parameters should be taken into account for the best choice of a MIMO-OFDM scheme.

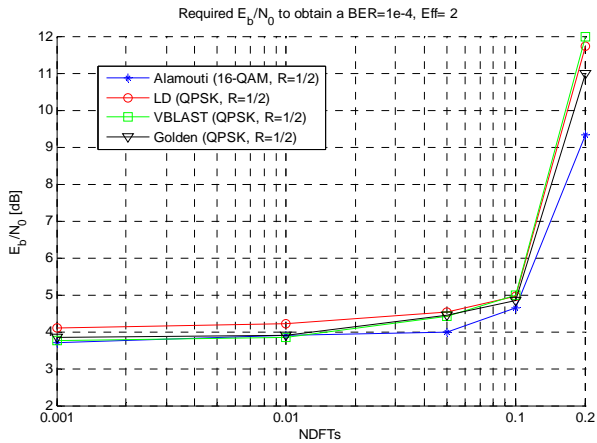

Figure 4- Required Eb/N0 to obtain a BER=10-4, Spectral efficiency η=2 [b/s/Hz], results obtained after 3 iterations for LD, VBLAST and Golden code.

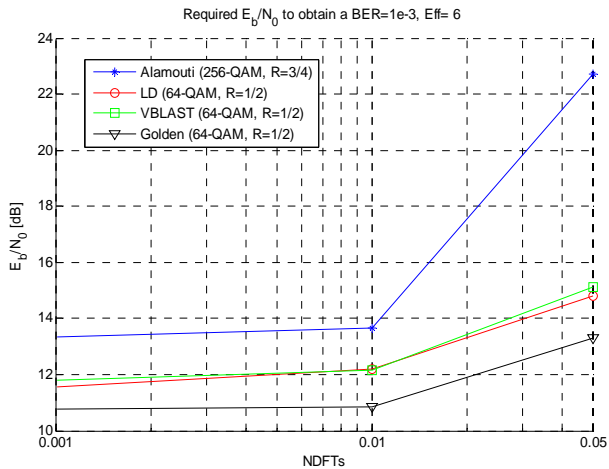

Figure 5- Required Eb/N0 to obtain a BER=10-4, Spectral efficiency η=6 [b/s/Hz], results obtained after 3 iterations for LD, VBLAST and Golden code.

## V. CONCLUSION

In this paper, we have investigated the effect of CFO on different MIMO-OFDM schemes for the second generation of terrestrial digital video broadcasting (DVB-T2). We showed that, for high spectral efficiency, the Alamouti scheme is more sensitive to CFO when compared with other NO schemes and the Golden code presents the best results.

## ACKNOWLEDGMENTS

The authors would like to thank the European CELTIC project "B21C" for its support of this work.